\documentclass[aps,preprint]{revtex4}%
\usepackage{amsfonts}
\usepackage{amsmath}
\usepackage{amssymb}
\usepackage{graphicx}%
\begin{document}
\title{The information loss problem and Hawking radiation as tunneling}
\author{Baocheng Zhang$^{1}$}
\email{zhangbaocheng@cug.edu.cn}
\author{Christian Corda$^{2}$}
\email{cordac.galilei@gmail.com}
\author{Qing-yu Cai$^{3,4,5}$}
\email{qycai@hainanu.edu.cn}
\affiliation{$^{1}$School of Mathematics and Physics, China University of Geosciences,
Wuhan 430074, China}
\affiliation{$^{2}$SUNY Polytechnic Institute, Utica, NY 13502, USA}
\affiliation{$^{3}$Center for Theoretical Physics, Hainan University, Haikou, 570228, China}
\affiliation{$^{4}$School of Information and Communication Engineering, Hainan University,
Haikou, 570228, China}
\affiliation{$^{5}$Peng Huanwu Center for Fundamental Theory, Hefei, 230026, Anhui, China}

\begin{abstract}
In this paper, we review some methods that tried to solve the information loss
problem. In particular, we revisit the solution based on Hawking radiation as
tunneling, and provide a detailed statistical interpretation on the black hole
entropy in terms of the quantum tunneling probability of Hawking radiation
from the black hole. In addition, we show that black hole evaporation is
governed by a time-dependent Schr\"{o}dinger equation that sends pure states
into pure states rather than into mixed states (Hawking had originally
established that the final result would be mixed states). This is further
confirmation of the fact that black hole evaporation is unitary.

\end{abstract}
\maketitle


\section{Introduction}

Due to Stephen Hawking's work, it became known that when general relativity is
combined with quantum theory, black holes can emit thermal radiation, which is
known as Hawking radiation \cite{swh74,swh75}. Shortly afterward, Hawking
discovered that regardless of the initial state that forms the black hole, it
will evolve into a thermal radiation state or a mixed state. This many-to-one
evolution not only violates the principle of unitarity in quantum mechanics
but also leads to the loss of information about the black hole's initial
state, which is known as the famous \textquotedblleft black hole information
loss paradox\textquotedblright\ \cite{swh76}. Since the discovery of the
\textquotedblleft black hole information loss paradox\textquotedblright, many
scientists have studied the issue, but so far, no solution has completely
resolved the problem. This is partly due to the lack of a complete theory of
quantum gravity, and partly because even within the framework of
semi-classical gravity, the existing solutions still face insurmountable or
unresolved issues. Past solutions can generally be categorized into the
following several approaches.

The first perspective argues that information is fundamentally lost. The main
proponent of this view was Hawking, but in 2005, he published an article
stating that information is not lost (\textquotedblleft elementary quantum
gravity interactions do not lose information or quantum
coherence\textquotedblright) \cite{swh05}. However, since no specific analysis
or calculation process was provided, this claim was not widely accepted.
Recently, Unruh and Wald \cite{uw17} reanalyzed and summarized this viewpoint,
arguing that the evolution from a pure state to a mixed state during the
process of black hole collapse and evaporation would not have any impact on
existing physics, so information loss is acceptable. However, this analysis of
the consequences remains speculative and does not fundamentally demonstrate
that information is indeed lost. Whether a true theory of quantum gravity
still suggests information loss remains unclear.

The second perspective suggests that information is still conserved in
semi-classical theory. Semi-classical theory refers to using classical theory
to describe gravity while using quantum theory for everything else. The most
straightforward view in this regard is that the problem of information loss
should not exist at all, as Peres and Terno stated at the end of their review
(\textquotedblleft There is no issue of information loss at
all\textquotedblright) \cite{pt04}. However, their analysis of the events seen
by different observers is not detailed enough to resolve all the potential
paradoxes revealed by the information loss problem.

An important recent development in the international study of the black hole
information loss problem is the work of scientists like Almheiri, Engelhardt,
and Penington \cite{aem19,gp20}, who used the AdS/CFT approach to find a way
to calculate the entropy of Hawking radiation in a class of special black
holes, proving that the entropy of Hawking radiation during the evaporation
process follows the Page curve. This is indeed an elegant solution, and this
approach has recently garnered significant attention, as noted in related
discussions \cite{jm20,att21}. However, upon closer analysis, it becomes clear
that this approach is still essentially semi-classical in nature. While this
solution is mathematically impressive, the physical shortcomings are also
evident. The primary issue is that it relies on analyzing quantum extremal
surfaces in higher-dimensional spacetime, and whether this extra dimension
actually exists remains unclear. Setting aside the calculation method itself,
it is still unclear how exactly information is encoded within the black hole
and how it emerges from the black hole. Of course, this is a common flaw in
all attempts to resolve the black hole information loss problem within the
framework of semi-classical theory.

The third perspective considers possible effects beyond current understanding
to solve the black hole information loss problem. In this regard, some
scientists suggest that black holes do not completely evaporate and may leave
behind remnants or transform into a \textquotedblleft baby
universe\textquotedblright\ \cite{acn87} to preserve the relevant information.
This idea has some merit and remains an area of ongoing research, as noted in
recent review articles \cite{coy15}.

Other scientists propose that black holes may possess not only mass, charge,
and angular momentum but also additional degrees of freedom on their surface
known as \textquotedblleft quantum hair\textquotedblright, which could store
information \cite{kw89}. However, if this method is used to solve the black
hole information loss paradox, a mechanism would be needed to transition
quantum gravity theory to low-energy local quantum field theory, and this
mechanism is not yet fully understood. Notably, the recent study by Hawking
and others that black holes possess \textquotedblleft soft
hair\textquotedblright\ could potentially aid progress on this issue
\cite{hps16}.

Another intriguing approach involves quantum information theory, specifically
related to quantum teleportation. However, the method known as
\textquotedblleft final state projection\textquotedblright\ \cite{hm04}, which
is associated with this concept, does not explain why or how a projection
occurs at the final stage of black hole evolution. A more radical idea is the
so-called \textquotedblleft firewall hypothesis\textquotedblright%
\ \cite{amps13}, which maintains the principle of unitarity in quantum
mechanics by violating the equivalence principle. Nevertheless, the firewall
itself presents a dilemma: either it truly exists, or an alternative solution
(see the \textquotedblleft ER=EPR\textquotedblright\ conjecture \cite{ms13})
must be found. For further solutions and related discussions, see recent
reviews \cite{jp16,dh16,dm17}.

All those methods mentioned above are based on the thermal spectrum discovered
by Hawking, but recently, Parikh and Wilczek \cite{pw00} developed a method of
Hawking radiation as tunneling with the energy conservation considered, which
broke down the fixed spacetime background and included the reaction in their
method, and thus the result of black hole radiation was gotten as non-thermal
naturally. Building on the non-thermal black hole radiation spectrum, it has
been demonstrated that this spectrum not only contains correlations that can
carry information but also that the total entropy (i.e., the sum of the
entropy of the remaining black hole and the radiation) during the black hole
radiation process remains conserved \cite{zcyz09}. These results are entirely
consistent with the requirements of quantum mechanical unitarity, suggesting
that the non-thermal radiation process based on energy conservation
considerations may indeed be unitary \cite{zczy13}. The fundamental physical
picture in this solution is that for black holes with the same mass but
different initial states, their radiation processes differ. The number of
different radiation processes can reveal the number of microscopic states
contained within the black hole. We used standard statistical methods to
demonstrate that this number of microscopic states can explain the area
entropy of the black hole.

Whether it is the entanglement entropy of the radiation process satisfying the
Page curve or the total entropy conservation of the black hole and radiation,
both only demonstrate that the radiation process is unitary. It seems
challenging to go further. This suggests that new elements need to be
introduced in future analyses. As previously discussed, the issue of unitarity
also involves a microscopic understanding of black hole entropy. In this
paper, we will briefly review how Hawking radiation, viewed as a tunneling
process, can demonstrate that the entire radiation process is unitary and
provide a microscopic explanation of black hole entropy based on the
probabilistic nature of the radiation process itself.

\section{Hawking radiation as tunneling}

Here we will review the mechanism to relate the reaction in the black hole
evaporation with particle annihilation. The reaction considered firstly in the
calculation of black hole radiation is derived from the tunneling method of
Parikh and Wilczek \cite{pw00}. In their method, they transformed the
Schwarzschild coordinates
\begin{equation}
ds^{2}=-\left(  1-\frac{2M}{r}\right)  dt^{2}+\left(  1-\frac{2M}{r}\right)
^{-1}dr^{2}+r^{2}d\Omega^{2}%
\end{equation}
into the Painlev\'{e} coordinates
\begin{equation}
ds^{2}=-\left(  1-\frac{2M}{r}\right)  dt^{2}+2\sqrt{\frac{2M}{r}}%
dtdr+dr^{2}+r^{2}d\Omega^{2},
\end{equation}
which is non-singular at the horizon $r=2M$. Then one could consider that the
tunneling of particles as emission impetus and calculate the probability as%
\begin{equation}
P\sim\exp[-8\pi E\left(  M-\frac{E}{2}\right)  ]=\exp\left(  \Delta S\right)
\label{nt}%
\end{equation}
where $E$ is the energy of tunneling particle and $S=4\pi M^{2}$ is the black
hole's Bekenstein-Hawking entropy. From Eq. (\ref{nt}), it is clear that black
hole radiation is a non-thermal spectrum, which differs from the thermal
spectrum originally obtained by Hawking. The main reason for this difference
is that, in Hawking's calculation of black hole radiation, the reaction effect
or the energy conservation was not taken into account. The spacetime
background he used was fixed, and thus his calculation may not have been
entirely rigorous. The result of thermal radiation he derived might be
corrected by considering the reaction effect. In fact, this is indeed the
case. When the reaction effect is considered, Parikh and Wilczek recalculated
the black hole radiation and found that the final radiation spectrum is not
purely thermal. Their calculation is based on the quantum mechanical tunneling
effect, where vacuum field fluctuations outside the black hole event horizon
generate particle-antiparticle pairs. The antiparticles, or negative energy
particles, will tunnel into the black hole, while the positive energy
particles may escape the black hole, forming black hole radiation.  The
tunneling barrier is caused by the reduction in the black hole's mass or by
the tunneling particles themselves. When the black hole mass is large, this
non-thermal spectrum can be approximated as a thermal spectrum and yields the
same temperature as Hawking's result. However, when the black hole mass is not
very large, the non-thermal nature of the spectrum becomes evident.

Since this spectrum is non-thermal, it is natural for us to ask whether there
is some correlation hidden in this radiation spectrum. This leads to the
question of how to determine whether a correlation exists. In statistics, we
determine if two events are correlated by checking whether the probability of
each event occurring individually is equal to the probability of both events
occurring simultaneously. If they are equal, then the two events are
statistically independent; otherwise, there must be a correlation between
them. We can understand this correlation in a simpler ways. For example, if
there are $m$ black balls and $n$ white balls in a box, and if we draw a black
ball first and then put it back, and then draw a white ball, it's clear that
these two events are completely independent. However, if we do not put the
black ball back after the first draw, then there is some correlation between
the two events. In the context of black holes, after the first particle is
radiated, it will certainly not fly back into the black hole. Therefore, the
second particle that is radiated will definitely have some correlation with
the first particle. Standard statistical analysis shows that such a
correlation exists in the non-thermal spectrum given in Eq. (\ref{nt}). So why
is there no such correlation in the thermal spectrum? This is because, when
deriving the thermal spectrum, the recoil effect is not considered. This is
equivalent to assuming that after the first particle is radiated, it flies
back into the black hole before the second particle is radiated. Clearly, in
this case, there would be no correlation between the two radiation events.

Now we consider a subsequent emission. At first, we will investigate whether
there are correlations existed between the two emissions with energies $E_{1}$
and $E_{2}$ respectively. When the first emission with the energy $E_{1}$
finishes, the tunneling probability for a particle of energy $E_{2}$ has to be
treated carefully, since the correlation can be assumed in advance. According
to the statistic theory described in the last paragraph, the two probabilities
can be obtained by taking the integral of their joint probability $P\left(
E_{1},E_{2}\right)  $, i.e. $P\left(  E_{1}\right)  =\int P\left(  E_{1}%
,E_{2}\right)  dE_{2}$ and $P\left(  E_{2}\right)  =\int P\left(  E_{1}%
,E_{2}\right)  dE_{1}$, where $P\left(  E_{1},E_{2}\right)  =P\left(
E_{1}+E_{2}\right)  $ is the joint probability of the two emissions with
energies $E_{1}$ and $E_{2}$ occuring simultaneously. Thus, we can confirm
that the correlation exists in the non-thermal radiation spectrum by finding
the conditional probability $P\left(  E_{2}|E_{1}\right)  =\frac{P\left(
E_{1},E_{2}\right)  }{P\left(  E_{1}\right)  }\neq P\left(  E_{2}\right)  $,
or by $P\left(  E_{1},E_{2}\right)  \neq P\left(  E_{1}\right)  \cdot P\left(
E_{2}\right)  $. From the perspective of the statistics, the correlation can
be measured as \cite{amv05},%
\begin{equation}
C\left(  E_{1}+E_{2};E_{1},E_{2}\right)  =\ln P\left(  E_{1},E_{2}\right)
-\ln\left[  P\left(  E_{1}\right)  \cdot P\left(  E_{2}\right)  \right]  .
\label{cq}%
\end{equation}

Since the correlation exists, we will continue to ask: can this correlation
carry information? Can it carry all the information? Our answer is yes. This
question involves the concept of mutual information in quantum information
theory. Mutual information describes the amount of information that can be
shared between two correlated events. Thus, it requires to calculate that the
correlation $C$ is equal to the mutual information. For a general composite
quantum system composed of subsystems $A$ and $B$, the mutual information is
defined as $S(A:B)\equiv S(A)+S(B)-S(A,B)=S(A)-S(A|B)$, where $S(A|B)$ is the
conditional entropy \cite{nc00}. For the situation of black hole tunneling
radiation, the entropy for the tunneling emitted particle is obtained as
$S\left(  E_{i}|E_{1},E_{2},\ldots,E_{i-1}\right)  =-\ln P(E_{i}|E_{1}%
,E_{2},\ldots E_{i-1})$, where $E_{i}$ is the energy of the tunneling particle
after the black hole has emitted particles with a total energy $E_{f}%
=\overset{i-1}{\underset{j=1}{\sum}}E_{j}$. Obviously, this is a conditional
entropy that measures the entropy of emission $E_{i}$ given that the values of
all the emitted particles with energies $E_{1},E_{2}$, $\ldots$, and $E_{i-1}$
are known. Through the calculation, it is not hard to find that the entropy
for the remaining black hole with mass $M-E_{f}$ decreases compared with the
initial entropy of the black hole, because the emitted particles carry
entropies. This balances the total entropy of the black hole and the
radiation, and will not lead to any violation of the thermodynamic second law
for a black hole \cite{swh760}. When mutual information is applied to the
emissions of two particles with energies $E_{1}$ and $E_{2}$, we have%
\begin{equation}
S(E_{2}:E_{1})\equiv S(E_{2})-S(E_{2}|E_{1})=-\ln P(E_{2})+\ln P(E_{2}|E_{1}).
\end{equation}
Then it is found that $S(E_{2}:E_{1})=8\pi E_{1}E_{2}$, which answers the
questions raised at the beginning of this paragraph. Alternatively, we can
understand this from another perspective: the lack of information manifests as
an increase in uncertainty about an event. In other words, because we do not
know the exact mechanical process when a coin is tossed, we cannot accurately
predict which side will land face up. However, we all believe that the
information about which side will land up is certainly embedded in some
complex process. For the black hole radiation process, the correlation we have
discovered will ultimately counteract the increased uncertainty due to the
lack of information, leading to the possibility that the entire process is unitary.

The discovery of the correlation provides us with a channel through which we
can understand the information leakage of black holes. However, there is
another question: does entropy remain non-increasing? In information theory,
entropy is a measure of uncertainty, but this measure cannot be directly
linked to correlation. Therefore, we must examine whether the entropy of the
black hole is conserved throughout the entire process, which is the essential
reauirement for the unitarity of the quantum mechanics. According to the above
analysis, except for the first radiation, the other radiations occur in the
form of conditional probabilities. Therefore, the entropy carried away by them
is also conditional entropy. In this way, we can calculate the total entropy
of black hole radiation as
\begin{equation}
S(E_{1},E_{2},\cdots,E_{n})=\sum\limits_{i=1}^{n}S(E_{i}|E_{1},E_{2}%
,\cdots,E_{i-1}), \label{te}%
\end{equation}
where $M=\sum_{i=1}^{n}E_{i}$ corresponds to the energy of all emissions due
to the energy conservation. Calculate the summation in Eq. (\ref{te}), and we
obtain that $S(E_{1},E_{2},...,E_{n})=4\pi M^{2}$ which is just the black
hole's Bekenstein-Hawking entropy. Our calculation shows that the continuous
tunneling black hole radiation process is an entropy-conserving process. Thus,
we can say that the black hole radiation process is a unitary process, and
information is not lost.

In particular, our analysis is still phenomenological and does not involve any
microscopic mechanisms. However, our analysis is meaningful, as it applies to
different types of black hole radiation. On one hand, it indicates that energy
and momentum conservation is key to solving the information loss problem. More
importantly, it also suggests that regardless of the microscopic mechanism,
the spectrum ultimately obtained should be this non-thermal spectrum;
otherwise, it could lead to the violation of information conservation.

\section{Statistic interpretation of entropy}

Since the radiation process is unitary, it is significant to understand the
black hole entropy based on Hawking radiation as tunneling. This can be
understood by counting the number of ways the black hole emits radiations
\cite{zcyz09,iy10,csy16}. The so-called ways refer to the following: the black
hole first radiates a particle with energy $E_{1}$, then it radiates a
particle with energy $E_{2}$, followed by a particle with energy $E_{3}$, and
so on, until the black hole has radiated all its energy. Actually, the energy
of the particle radiated in the first event could be $E_{1}^{^{\prime}}\neq
E_{1}$, the energy of the particle radiated in the second event could be
$E_{2}^{^{\prime}}\neq E_{2}$, and so on, until the black hole has radiated
all its energy. This is a different radiation way compared to the former. Note
that as long as there is one difference for the energy of emitted particles,
it can be defined as a new radiation way.

We denote each radiation way as a microstate $\left(  E_{1},E_{2},\cdots
,E_{n}\right)  $ and $\sum_{i}E_{i}=M$. It is not hard to obtain the
probability for the microstate $\left(  E_{1},E_{2},\cdots,E_{n}\right)  $ as
$P_{t}=P(E_{1})\ast P(E_{2})\ast\cdots\ast P(E_{n})$ with $P(E_{1})=\exp(-8\pi
E_{1}(M-E_{1}/2))$, $P(E_{2})=\exp(-8\pi E_{2}(M-E_{1}-E_{2}/2))$, $\cdots$,
$P(E_{n})=\exp(-8\pi E_{n}(M-E_{1}-E_{2}-\cdots-E_{n-1},E_{n}/2))=\exp(-4\pi
E_{n}^{2})$. After a detailed calculation, we can obtain that
\begin{equation}
P_{t}=\exp(-4\pi M^{2})=\exp(-S_{BH}),
\end{equation}
where $S_{BH}$ is the entropy of black hole. Further, we can obtain the number
of the microstates as $\Omega=\frac{1}{P_{t}}=\exp(S_{BH})$, according to the
fundamental postulate of statistical mechanics that all microstates of an
isolated system are equally likely. This provides an feasible interpretation
for the Bekenstein-Hawking entropy $S_{BH}$, that is $S=\ln\Omega=S_{BH}$, in
terms of the number of ways for evaporation\textbf{ }according to the
Boltzmann's definition. On one hand, this provides a statistically microscopic
explanation for the entropy of a black hole; on the other hand, it also shows
that black hole radiation can carry away all the entropy of the black hole,
satisfying the requirement of unitarity in quantum mechanics.

\section{Time-dependent Schr\"{o}dinger equation for black hole evaporation}

The fact that successive emissions of Hawking particles are countable, as it
has been shown in the previous Sections, is consistent with what was
originally found by Bekenstein in 1974 \cite{Bekenstein}, that the energy
spectrum of a black hole is discrete. Bekenstein indeed obtained $E_{n}%
=\sqrt{\frac{n}{2}}$ by using the Bohr-Sommerfeld quantization condition
because he argued that the Schwarzschild black hole behaves as an adiabatic
invariant. A similar result was found in \cite{Corda}, starting from the
quantization of the famous Oppenheimer-Snyder gravitational collapse:
\begin{equation}
E_{n}=-\sqrt{\frac{n}{4}}. \label{eq: BH energy levels finale.}%
\end{equation}
In quantum mechanics, time evolution of perturbations can be described by an
operator \cite{Corda}

\emph{
\begin{equation}
U(t)=%
\begin{array}
[c]{c}%
W(t)\;\;\;for\;0\leq t\leq\tau\\
0\;\;\;for\;t<0\;and\;t>\tau.
\end{array}
\label{eq: perturbazione}%
\end{equation}
}Then, the complete (time dependent) Hamiltonian is described by the operator
\cite{Corda}
\begin{equation}
H(r,t)\equiv V(r)+U(t), \label{eq: Hamiltoniana completa}%
\end{equation}
where $V(r)$ is given by Eq. (48) in \cite{Corda}. Thus, considering a wave
function $\psi(r,t),$ we can write the correspondent \emph{time dependent
Schr\"{o}dinger equation }for the system (i.e. the evaporating black hole)as
\cite{Corda}
\begin{equation}
i\frac{d|\psi(r,t)>}{dt}=\left[  V(r)+U(t)\right]  |\psi(r,t)>=H(r,t)|\psi
(r,t)>. \label{eq: Schroedinger equation}%
\end{equation}
The\emph{ }state\emph{ }which satisfies Eq. (\ref{eq: Schroedinger equation})
is \cite{Corda}
\begin{equation}
|\psi(r,t)>=\sum_{n}a_{n}(t)\exp\left(  -iE_{n}t\right)  |\varphi_{n}(r)>,
\label{eq: Schroedinger wave-function}%
\end{equation}
where the $\varphi_{n}(r)$ are the eigenfunctions of the time independent
Schr\"{o}dinger equation in Eq. (49) in \cite{Corda} and the $E_{n}$ are the
correspondent eigenvalues. Now, we closely follows \cite{Corda}. In the basis
$|\varphi_{n}(r)>$, the matrix elements of $W(t)$ can be written as%

\begin{equation}
W_{ij}(t)\equiv A_{ij}\delta(t), \label{eq: a delta}%
\end{equation}
where $W_{ij}(t)=<\varphi_{i}(r)|W(t)|\varphi_{j}(r)>$ and the $A_{ij}$ are
real. In order to solve the complete quantum mechanical problem described by
the operator (\ref{eq: Hamiltoniana completa}), we need to find the
probability amplitudes $a_{n}(t)$ due to the application of the perturbation
described by the time dependent operator (\ref{eq: perturbazione}), which
represents the perturbation associated to the emission of a Hawking particle.
For $t<0,$ i.e. before the perturbation operator (\ref{eq: perturbazione})
starts to work, the system is in a stationary state $|\varphi_{m}(t,r)>,$ at
the quantum level $m,$ with energy $E_{m}=-\frac{1}{2}\sqrt{m},$ given by Eq.
(\ref{eq: BH energy levels finale.}). Thus, in Eq.
(\ref{eq: Schroedinger wave-function}) only the term%

\begin{equation}
|\psi_{m}(r,t)>=\exp\left(  -iE_{m}t\right)  |\varphi_{m}(r)>,
\label{eq: Schroedinger wave-function in.}%
\end{equation}
is not null for $t<0.$ This implies $a_{n}(t)=\delta_{nm}\:\:$for $\:t<0.$
When the perturbation operator (\ref{eq: perturbazione}) stops to work, i.e.
after the emission, for $t>\tau$ the probability amplitudes $a_{n}(t)$ return
to be time independent, having the value $a_{m\rightarrow n}(\tau)$. In other
words, for $t>\tau\:$ the system is described by the\emph{ wave function
$\psi_{final}(r,t)$ }which corresponds to the state%

\begin{equation}
|\psi_{final}(r,t)>=\sum_{n=1}^{m}a_{m\rightarrow n}(\tau)\exp\left(
-iE_{n}t\right)  |\varphi_{n}(x)>. \label{eq: Schroedinger wave-function fin.}%
\end{equation}
Therefore, the probability to find the system in an eigenstate having energy
$E_{n}=-\sqrt{n}$, with $n<m$ for emissions, is given by%

\begin{equation}
\Gamma_{m\rightarrow n}(\tau)=|a_{m\rightarrow n}(\tau)|^{2}.
\label{eq: ampiezza e probability}%
\end{equation}
By using a standard analysis, we obtain the following differential equation
from Eq. (\ref{eq: Schroedinger wave-function fin.})%

\begin{equation}
i\frac{d}{dt}a_{m\rightarrow n}(t)=\sum_{l=1}^{n}W_{ml}a_{m\rightarrow
l}(t)\exp\left[  i\left(  \Delta E_{l\rightarrow n}\right)  t\right]  .
\label{eq: systema differenziale}%
\end{equation}
To first order in $U(t)$, by using the Dyson series \cite{Corda}, the solution
is obtained as%

\begin{equation}
a_{m\rightarrow n}=-i\int_{0}^{t}\left\{  W_{nm}(t^{\prime})\exp\left[
i\left(  \Delta E_{m\rightarrow n}\right)  t^{\prime}\right]  \right\}
dt^{\prime}. \label{eq: solution}%
\end{equation}
Now, we insert Eq. (\ref{eq: a delta}) in Eq. (\ref{eq: solution}) obtaining
\begin{equation}
a_{m\rightarrow n}=iA_{nm}\int_{0}^{t}\left\{  \delta(t^{\prime})\exp\left[
i\left(  \Delta E_{m\rightarrow n}\right)  t^{\prime}\right]  \right\}
dt^{\prime}=\frac{i}{2}A_{nm}. \label{eq: solution 2}%
\end{equation}
This equation can be combined with Eq. (144) in \cite{Corda} and with Eq.
(\ref{eq: ampiezza e probability}). We obtain%

\begin{equation}%
\begin{array}
[c]{c}%
\alpha\exp\left[  16\pi\left(  n-m\right)  \right]  =\frac{1}{4}A_{nm}^{2}\\
\\
A_{nm}=2\sqrt{\alpha}\exp\left[  8\pi\left(  n-m\right)  \right] \\
\\
a_{m\rightarrow n}=-i\sqrt{\alpha}\exp\left[  8\pi\left(  n-m\right)  \right]
.
\end{array}
\label{eq: uguale}%
\end{equation}
As $\sqrt{\alpha}\sim1,$ we find $A_{nm}\sim10^{-11}$ for $n=m-1$, i.e. when
the probability of emission has its maximum value. This implies that second
order terms in $U(t)$ are $\sim10^{-22}$ and that we can, in turn, neglect
them. Clearly, for $n<m-1$, we get a better approximation, because the
$A_{mn}$ are even smaller than $10^{-11}$. Hence, we can write down the final
form of the ket representing the state as%

\begin{equation}
|\psi_{final}(r,t)>=\sum_{n=1}^{m}-i\sqrt{\alpha}\exp\left[  8\pi\left(
n-m\right)  -iE_{n}t\right]  |\varphi_{n}(r)>.
\label{eq: Schroedinger wave-function finalissima}%
\end{equation}
The\emph{ }state (\ref{eq: Schroedinger wave-function finalissima}) represents
a \emph{pure final state instead of a mixed final state.} Therefore, the
states are obtained in terms of an \emph{unitary} evolution matrix instead of
a density matrix\emph{ }and\emph{ }this confirms the fundamental conclusion
argued in previous Sections that\emph{ }information is not loss in black hole
evaporation. This result is consistent with 't Hooft's idea that
Schr\"{o}dinger equations can be used universally for all dynamics in the
universe \cite{Corda} and dismisses the claim of Hawking that the final result
of black hole evaporation would be mixed states \cite{swh76}. The final state
of Eq. (\ref{eq: Schroedinger wave-function finalissima}) is due to potential
arbitrary transitions $m\rightarrow n$, with $m>n$. Then, the subsequent
\emph{collapse of the wave function} to a new stationary state, at the quantum
level $n$
\begin{equation}
|\psi_{n}(r,t)>=\exp\left(  -iE_{n}t\right)  |\varphi_{n}(r)>,
\label{eq: Schroedinger wave-function out}%
\end{equation}
implies that the wave function of the infalling particle in Hawking's
mechanism of particles creation by black holes has been transferred to the
black hole excited state at the quantum level $n$ \cite{Corda} and it is given
by
\begin{equation}
|\psi_{\left(  m\rightarrow n\right)  }(r,t)>\equiv\exp\left(  -iE_{n}%
t\right)  |\varphi_{n}(r)>-\exp\left(  -iE_{m}t\right)  |\varphi_{m}(r)>.
\label{eq: funzione onda particella emessa}%
\end{equation}
This wave function results entangled with the wave function of the particle
which propagates towards infinity. Clearly, the evolution of black hole
evaporation that it has been discussed in this Section is \emph{unitary}.

\section{Discussion and Conclusion}

It is noted that the discussion about the information loss problem based on
our method above has been extended to many different types of black holes
\cite{shp10,zczy11-1,zczy11-2,zcz12,shp12,shm12,zczy14,ch15,ch15-2,cch19,dh19,sl23}
and many different situations
\cite{cs09,bcz09,bp11,zczy11-3,cc12,dp13,mf13,chk13,svc14,chs14,dcs14,cc15,cc15-2,cc15-3,yl16,sb17,gc18,pg22,tz23,xm24}
(also see the review paper \cite{vac11}). All those show that the Hawking
radiation as tunneling should be a unitary process.

However, when we consider that information can be carried out by Hawking
radiation, there is still another question that needs to be resolved: Where is
the information stored before it is carried out by Hawking radiation? Before
Hawking discovered black hole radiation, the renowned American general
relativity expert Wheeler provided an explanation for classical black holes
swallowing information: the information is stored inside the black
hole---although it is not lost, it is impossible for an external observer to
retrieve it. In this view, when the black hole does not radiate or has not yet
completed its radiation, people assume that the information remains stored
within the black hole. But can the information actually be inside the black
hole? This question is extremely difficult to answer. First, it is not clear
what exactly exists inside a black hole, but some studies on black hole volume
offer intriguing insights. In 2015, Christodoulou and Rovelli showed that a
black hole formed through collapse has an extremely large volume \cite{cr15}.
Whether there are enough degrees of freedom in such a large volume to store
information had been studied. It indicated that the degrees of freedom within
the black hole's volume can indeed statistically produce entropy proportional
to the black hole's surface area, but this entropy is much smaller than the
black hole's actual entropy \cite{bz15,bz17,by17}. Therefore, the black hole's
interior likely does not have enough degrees of freedom to store all the
information. However, this does not completely rule out the possibility that
information is stored inside the black hole \cite{zy20}.

As for the possibility that information is stored on the event horizon, this
is considered more likely, given that black hole entropy is proportional to
its surface area. Discussions involving area quantization and entanglement
entropy also support this possibility. In this context, the idea of
\textquotedblleft quantum hair\textquotedblright\ \cite{kw89}\ is worth
mentioning, particularly the recent conclusion by Hawking and others that
black holes possess \textquotedblleft soft hair\textquotedblright%
\ \cite{hps16}. Based on the idea of \textquotedblleft soft
particles\textquotedblright\ in quantum field theory and the BMS symmetry of
asymptotically flat spacetime, they found that \textquotedblleft soft
hair\textquotedblright\ exists in black hole spacetime and can carry
information. When this \textquotedblleft soft hair\textquotedblright\ on the
black hole's event horizon is excited, it could explain the black hole's area
entropy. However, whether this \textquotedblleft soft hair\textquotedblright%
\ carries information about the black hole's initial state or whether this
information can be retrieved by external observers remains unclear.

Another question is whether Hawking radiation truly has a non-thermal
spectrum, as suggested by Parikh and Wilczek. This may need to be answered
through observation or experimentation. In the context of astrophysics,
Hawking radiation is almost unobservable due to its extremely low temperature
(for example, the radiation temperature of a black hole with the mass of the
Sun is approximately $10^{-7}$ K, which is seven orders of magnitude lower
than the $2.7$ K temperature of the cosmic microwave background).

It should also be remembered a recent interesting approach, originally carried
out by Vaz in 2014 \cite{Vaz} and recently further developed by one of us (CC)
\cite{Corda3, Corda4}. In fact, in 2014 Hawking \cite{SWH} proposed that BH
event horizons could not be the final result of the gravitational collapse. He
speculated that the BH event horizon should be replaced by an
\textquotedblleft apparent horizon\textquotedblright\ where infalling matter
is suspended and then released. Hawking did not give a mechanism for how this
can work, which was later given by Vaz \cite{Vaz}, who supported Hawking's
conclusion. Vaz indeed discussed an interesting quantum gravitational model of
inhomogeneous dust collapse by showing that continued collapse to a
singularity can only be achieved by combining two independent and entire
solutions of the Wheeler-DeWitt equation \cite{Vaz}. He argued that such a
combination is forbidden leading in a natural way to matter condensing on the
a pparent horizon during quantum collapse, forming a thin and dense
spherically symmetric shell \cite{Vaz}. A similar result had already been
obtained by Einstein in 1939 \cite{Einstein}. In \cite{Corda3, Corda4}, it was
shown that these thin and dense spherically symmetric shells have quantum
properties and obey the Schrodinger equation in the non-relativistic case and
the Klein-Gordon equation in the relativistic case. Well-defined energy
spectra correspond to these equations. Furthermore, the equations themselves
were derived from the historical homogeneous Oppheneimer-Snyder gravitational
collapse. Nontrivial consequences emerge from these results: i) black holes
have neither horizons nor singularities; ii) there is neither information loss
in black hole evaporation, nor black hole complementarity, nor firewall
paradox. Thus, in this case, the information problem is, in principle, solved
by showing that BHs actually have a different physical structure, that is the
structure of normal bodies emitting radiation from their surface like any
other body. Something similar also happens in the so-called \textquotedblleft
fuzzball paradigm\textquotedblright\ \cite{MM}. Fuzzballs are objects
predicted by string theory, intended to provide a fully BH quantum
description. Recalling that Bekenstein entropy is obtained through the count
of brane bound states, the fuzzball construction of BH microstates implies
that these states have no horizon and radiate from their surface like a normal
body, so there is no information paradox \cite{MM}. In this framework, quantum
gravity effects modify the entire region inside the horizon. Thus, one finds a
fuzzball, that is a ball of stringy matter with no horizon. Fuzzball radiates
from their surface like normal bodies, that is, not by the creation of pairs
from the vacuum. Thus, one finds no information paradox. The fuzzball paradigm
arises from the discovery of D-branes and the BH construction in terms of
bound states of branes. An estimate of the radius of brane bound states gives
a result of order of the horizon radius. Several families of brane bound
states can be, in turn, constructed, by finding a \textquotedblleft
fuzzball\textquotedblright; i.e., an object with no horizon, see \cite{MM} for
details. What the Einstein-Vaz shells and fuzzballs have in common is that
quantum gravity corrections become necessary at the horizon scale rather than
at the Planck scale as is believed by most researchers. In other words,
quantum corrections to classical general relativity depend on an energy scale
rather than a distance scale.

Over the past 40 years, scientists have discovered that it is possible to
simulate analogue gravity \cite{csv11} in many physical systems. In recent
years, significant progress has been made on the experimental front,
especially with Bose-Einstein condensates, where Hawking radiation has
reportedly been observed \cite{ngs19}. Our previous analysis provided an
experimental signal \cite{zczy13-2} for Hawking radiation as a tunneling
process. We hope that future experiments can observe the exact form of Hawking
radiation in more detail, which would be of great significance in solving the
black hole information loss problem.

\section{Acknowledgements}

This work is supported by National Natural Science Foundation of China (NSFC)
with Grant Nos. 12375057, 11947301, 12047502, and the Fundamental Research
Funds for the Central Universities, China University of Geosciences (Wuhan).

\end{document}